\documentclass{elsart}
\usepackage[dvips]{graphicx}
\usepackage{amsmath}

\newcommand{\kstar}{\ensuremath{K^{*}(892)^0}}
\newcommand{\khigh}{\ensuremath{K^{*}(1680)^0}}
\newcommand{\klow}{\ensuremath{K^{*}_{0}(1430)^0}}
\newcommand{\ktensor}{\ensuremath{K^{*}_{2}(1430)^0}}

\newcommand{\kstarb}{\ensuremath{\overline{K}{}^{*}(892)^0}}
\newcommand{\khighb}{\ensuremath{\overline{K}{}^{*}(1680)^0}}
\newcommand{\klowb}{\ensuremath{\overline{K}{}^{*}_{0}(1430)^0}}

\newcommand{\Dkstarmunu}{\ensuremath{D^+ \!\rightarrow\! \kstarb \mu^+ \nu}}
\newcommand{\Dkhighmunu}{\ensuremath{D^+ \!\rightarrow\! \khighb \mu^+ \nu}}
\newcommand{\Dklowmunu}{\ensuremath{D^+ \!\rightarrow\! \klowb \mu^+ \nu}}

\newcommand{\Dkpimunu}{\ensuremath{D^+ \!\rightarrow\! K^- \pi^+ \mu^+ \nu }}
\newcommand{\mkpi}{\ensuremath{m_{\textrm{K$\pi$}}}}

\newcommand{\thv}{\ensuremath{\theta_\textrm{v}}}
\newcommand{\thl}{\ensuremath{\theta_\ell}}
\newcommand{\costhv}{\ensuremath{\cos{\thv}}}

\newcommand{\qsq}{\ensuremath{q^2}}

\newcommand{\prd}[1]{Phys.~Rev.~D \textbf{#1}}
\newcommand{\plb}[1]{Phys.~Lett.~B \textbf{#1}}
\newcommand{\prl}[1]{Phys.~Rev.~Lett. \textbf{#1}}
\newcommand{\zpc}[1]{Z.~Phys.~C \textbf{#1}}
\newcommand{\npb}[1]{Nucl.~Phys.~B \textbf{#1}}

\newcommand{\andree}[3]{{#1}\makebox[0pt][l]{\raisebox{1.5ex}{}\raisebox{1.5ex}{#2}}\raisebox{-1.5ex}{}\raisebox{-1.5ex}{#3}}

\newcommand{\mysection}[1]{}
\newcommand{\mysubsection}[1]{}
\newcounter{saveeqn}%

\begin{document}

\begin{frontmatter}
\title{{\bf Hadronic Mass Spectrum Analysis of \Dkpimunu{} Decay and Measurement of the \kstar{} Mass and Width}}
\collaboration{The FOCUS Collaboration}

\author[ucd]{J.~M.~Link}
\author[ucd]{P.~M.~Yager}
\author[cbpf]{J.~C.~Anjos}
\author[cbpf]{I.~Bediaga}
\author[cbpf]{C.~G\"obel}
\author[cbpf]{A.~A.~Machado}
\author[cbpf]{J.~Magnin}
\author[cbpf]{A.~Massafferri}
\author[cbpf]{J.~M.~de~Miranda}
\author[cbpf]{I.~M.~Pepe}
% ...

\address[ucd]{University of California, Davis, CA 95616}
\address[cbpf]{Centro Brasileiro de Pesquisas F\'{\i}sicas, Rio de Janeiro, RJ, Brasil}
% ...

\nobreak
\begin{abstract}

We present a $K\pi$ mass spectrum analysis of the four-body semileptonic charm decay \Dkpimunu{} in the range of 0.65 GeV/$c^2$ $< \mkpi <$ 1.5 GeV/$c^2$. We observe a non-resonant contribution of $5.30 \pm 0.74^{\,+\,0.99}_{\,-\,0.96}\%$ with respect to the total \Dkpimunu{} decay. For the \kstar{} resonance, we obtain a mass of $895.41 \pm 0.32^{\,+\,0.35}_{\,-\,0.43}$ MeV/$c^2$, a width of $47.79 \pm 0.86^{\,+\,1.32}_{\,-\,1.06}$ MeV/$c^2$, and a Blatt-Weisskopf damping factor parameter of $3.96 \pm 0.54^{\,+\,1.31}_{\,-\,0.90}$ GeV$^{-1}$. We also report 90\% CL upper limits of 4\% and 0.64\% for the branching ratios $\frac {\Gamma (D^+ \to \khighb \mu^+ \nu)} {\Gamma (D^+ \to K^- \pi^+ \mu^+ \nu)}$ and $\frac{\Gamma (D^+ \to \klowb \mu^+ \nu)} {\Gamma (D^+ \to K^- \pi^+ \mu^+ \nu)}$, respectively.

\end{abstract}
\end{frontmatter}

\newpage

\mysection{Introduction}

Weak semileptonic decays of charm mesons continue to attract interest due to the relative simplicity of their theoretical description: the matrix element of these decays can be factorized as the product of the leptonic and hadronic currents. This makes the \Dkpimunu{} decay a natural place to study the $K\pi$ system in the absence of interactions with other hadrons. Due to Watson's Theorem~\cite{bigi,watson}, the observed $K\pi$ phase shifts in $D^+\!\rightarrow\! K^-\pi^+\mu^+\nu$ should be the same as those measured in $K\pi$ elastic scattering. 

It is known that the $K\pi$ final state of \Dkpimunu{} decay is strongly dominated by the \kstar{} vector resonance \cite{kstardominance,e687}. The large and clean sample of $D^+\!\rightarrow\! K^-\pi^+\mu^+\nu$ events collected by the Fermilab FOCUS experiment provides an excellent opportunity to measure the \kstar{} mass and width, as well as the effective Blatt-Weisskopf damping factor parameter discussed in Ref. \cite{blatt}. We also search for structures other than the \kstar{} resonance in the mass range of 0.65 GeV/$c^2$ $< \mkpi{} <$ 1.5 GeV/$c^2$.

The first suggestion that the \Dkpimunu{} decay proceeds via states other than the \kstar{} resonance comes from the Fermilab E687 experiment~\cite{e687}. The presence of an additional structure was confirmed by FOCUS in the analysis of the angular decay distributions, in which the \kstar{} form factor was measured ~\cite{topher1,topher2}. Specifically, significant discrepancies were found between the data and the predicted \Dkstarmunu{} angular decay distributions. A nearly constant amplitude and phase contribution to the helicity zero amplitude of the virtual $W^+$ was required to adequately fit the observed decay angular distributions.  The s-wave amplitude, $a_0 e^{i \delta_0}$, was measured in the vicinity of the \kstar{} pole with parameters $a_0 = 0.330 \pm 0.022 \pm 0.015$ and $\delta_0 = 0.68 \pm 0.07 \pm 0.05$. This new component accounts for 5\% of the \Dkpimunu{} branching fraction.

Motivated by this earlier FOCUS result, we search for other contributions in the $K\pi$ spectra. Specifically, we look for a possible contribution from the \khigh{}, \klow{}, and $\kappa$. We also present a more complete description of the non-resonant contribution. The existence of the $\kappa$, reported in~\cite{dkpipi}, remains controversial due to difficulties in the theoretical treatment of broad scalar states and the absence of a clear observation of this state in scattering experiments. Many models predicting the decay width of semileptonic decays, such as ISGW2~\cite{isgur} and QCD Sum-Rules~\cite{dosch}, indicate the tendency for these decays to proceed via low mass structures. In~\cite{marina} it is suggested that if the $\kappa$ has a substantial $q \bar{q}$ component in its wave function, it could account for more than 10\% of the \Dkpimunu{} decay rate.

\mysection{Experimental and analysis details}

The data were collected in the Wideband photoproduction experiment FOCUS during the Fermilab 1996--1997 fixed-target run. In FOCUS, a forward multi-particle spectrometer is used to measure the interactions of high energy photons on a segmented BeO target. The FOCUS detector is a large aperture, fixed-target spectrometer with excellent vertexing and particle identification. The FOCUS beamline~\cite{beam} and detector~\cite{topher1,detector1,detector2,detector3} have been described elsewhere.

To isolate \Dkpimunu{} events, we require that the muon, pion, and kaon candidate tracks have a 5\% or greater confidence level to originate from a common secondary vertex. Background is reduced by requiring the secondary decay vertex be separated from the production (primary) vertex by greater than $10\,\sigma_\ell$, where $\sigma_\ell$ is the uncertainty on the separation between the primary and secondary vertices. Possible backgrounds from higher multiplicity charm decays are suppressed by requiring the $K^-\pi^+\mu^+$ vertex be isolated from other tracks in the event (excluding tracks from the primary vertex). Specifically, we require that the maximum confidence level for another track to form a vertex with the secondary vertex candidate be less than 1\%. To suppress background from secondary interactions the decay vertex candidate must lie outside any target foil or detector material.

The muon, pion, and kaon candidates are selected in the following way. The muon track must have hits in at least 5 of the 6 segmented scintillator layers which comprise the inner muon detector and a muon confidence level exceeding 5\% (based on the fit to the hits).  The pion and kaon tracks must have a muon confidence level less than 0.1\%. The kaon is required to have a \v Cerenkov light pattern more consistent with that of a kaon than that of a pion by 2 units of log-likelihood, while the pion track is required to have a light pattern favoring the pion hypothesis over that of the kaon by 2 units~\cite{detector2}. In addition, the pions and kaons are required to have momenta greater than 5 GeV/$c$, while the muon momentum must exceed 10 GeV/$c$.

To suppress background from $D^+\!\rightarrow\! K^-\pi^+\pi^+$, we require that the invariant mass of the three tracks, where the muon candidate is given the pion mass, is less than 1.8 GeV/$c^2$. To suppress background from $D^{*+} \!\rightarrow\! D^0 \pi^+ \!\rightarrow\! (K^-\mu^+\nu) \pi^+$ and $D^{*+} \!\rightarrow\! D^0 \pi^+ \!\rightarrow\! (K^{*-}\mu^+\nu) \pi^+$, we require $m(K^-\mu^+\pi^+)-m(K^-\mu^+) > 0.2$ GeV/$c^2$. A total of 18245 \Dkpimunu{} candidates remain after the selection criteria.

The charm background, charm decays that are not \Dkpimunu{}, is estimated from more than one billion charm Monte Carlo events that pass through the entire data analysis chain. Our Monte Carlo is based on \textsc{Pythia}~\cite{pythia} and incorporates all known charm decays. The charm Monte Carlo sample was scaled to the data sample size using the fitted yield of the $D^+\!\rightarrow\! K^-\pi^+\pi^+$ signal. To estimate the background contribution coming from non-charm events, we define a wrong sign sample (WS) formed by $K^+\pi^-\mu^+$ tracks in the secondary vertex. We assume that non-charm events populate the wrong sign and right sign (RS) samples equally. The non-charm background distribution is obtained by subtracting the WS charm background (obtained from the Monte Carlo sample) from the WS data sample. We estimate the charm and non-charm background contributions to be, respectively, 17.8\% and 3.2\% of the total number of events over our signal region.

\mysection{The \Dkpimunu{} signal description}

Four-body decays of spinless particles are described by five kinematic variables. The variables chosen in this analysis are the $K^-\pi^+$ invariant mass (\mkpi{}), the square of the $\mu^+\nu$ mass (\qsq{}), and three decay angles: the angle between the $\pi^+$ and the $D^+$ direction in the $K^-\pi^+$ rest frame (\thv{}), which defines one decay plane, the angle between the $\nu$ and the $D^+$ direction in the $\mu^+\nu$ rest frame (\thl{}), which defines the second decay plane, and the acoplanarity angle ($\chi$) between these two decay planes. 

The differential decay rate can be represented by a coherent sum of resonant and non-resonant contributions to the angular momentum eigenstates of the $K^-\pi^+$ system, 
\begin{equation}
\frac{d\Gamma}{d m_{K\pi}} = \int {\left| \sum_{J} \sum_{R} a_{J,(R)} ~\mathcal{M}_{J} ~\mathcal{A}_{J,(R)} \right|^2 }~\phi ~d\Omega
\label{eq1}
\end{equation}

where $d\Omega \equiv d\qsq~d\!\cos{\thv}~d\!\cos{\thl}~d\chi$, $\mathcal{M}_{J}$ is the weak matrix element for a transition with angular momentum $J$, $\mathcal{A}_{J,(R)}$ represents the form of the hadronic final state amplitude contribution of resonance $R$ (or non-resonant) with strength $a_{J,(R)}$, and $\phi$ is the phase space density.

The possible  resonant states that couple to $K^-\pi^+$ are the scalars $\kappa$ and \klow{}, the vectors \kstar{} and \khigh{}, and the tensor \ktensor{}. The non-resonant contribution is assumed to be scalar. \footnote{Although the simplest way to obtain the forward-backward asymmetry described in Ref.~\cite{topher1} is to assume an s-wave amplitude interfering with the $K^*(892)^0$ (as was done in Ref~\cite{topher1}), small spin 2 components cannot be excluded.} Small amplitude contributions are most likely to be observed through the interference with large amplitude components. Due to the orthogonality of states with different angular momentum, only amplitudes with the same spin will produce significant interference contributions to the \mkpi{} mass spectrum, given our reasonably uniform angular acceptance. Therefore, the small vector \khigh{} and scalar \klow{} contributions might produce an observable effect on the \mkpi{} spectrum through their interference with the \kstar{} and a low mass s-wave amplitude, respectively. By contrast, the inclusion of a small \ktensor{} resonance contribution is unlikely to be observed, since it is orthogonal to the (dominant) \kstar{} and low mass s-wave amplitudes. For this reason we do not include the \ktensor{} resonance in our fits to the \mkpi{} spectrum in \Dkpimunu{}.

The parametrization of resonant states with angular momentum $J$ is given by the product of a Breit-Wigner\footnote{The Breit-Wigner form used by FOCUS differs by a factor of $-1$ from the LASS~\cite{lass} form.} and the normalized $R \to K^-\pi^+$ coupling, $\mathcal{F}_{J}$

\begin{equation}
\mathcal{A}_{J,R} = \frac{m_0\,\Gamma_0}{m_{K\pi}^2 - m_0^2 + i\,m_0\,\Gamma(m_{K\pi})}\,\mathcal{F}_{J} 
(m_{K\pi})
\label{eq2}
\end{equation}

where $\Gamma(m_{K\pi}) = \Gamma_0\,\mathcal{F}_{J}^2\,\frac{p^*}{p^*_0}\,\frac{m_0}{m_{K\pi}}$, $p^{*}$ is the magnitude of the kaon momentum in the resonance rest frame, $p^{*}_0 = p^{*}(m_{0})$, $\mathcal{F}_0 = 1$, and $\mathcal{F}_1 = \frac{p^*}{p^*_0}\,\frac{B(p^{*})}{B(p^{*}_0)}$. $B$ is the Blatt-Weisskopf damping factor given by $B = 1 / \sqrt{1 + r_0^2 ~p^{*2}}$~\cite{blatt}. The damping factor adds an additional fit parameter, $r_0$, in our fits to the \kstar{} line shape. The line shape of the $\kappa$ resonance is expected to deviate significantly from a pure Breit-Wigner, due to its large width and the close vicinity of the $K\pi$ threshold. In this analysis we use the $\kappa$ line-shape adopted by E791~\cite{dkpipi}. 
 
We use an empirical parametrization from $K^-\pi^+$ elastic scattering experiments for the non-resonant amplitude. A partial wave analysis performed by LASS observed that the s-wave amplitude can be represented as the sum of a \klow{} resonance coupled to $K^- \pi^+$ and $K \eta'$ and a smooth shape, consistent with the non-resonant hypothesis~\cite{lass}. \footnote{Charm decays are traditionally fit to a model where the strengths of both resonant and non-resonant contributions are fit parameters. Hence we will independently adjust the non-resonant and \klow{} resonant contributions found by LASS to best fit our data.} LASS fitted the non-resonant component to an effective range model of the form
\begin{equation}
\cot{\delta_{\textrm{LASS}}} = \frac{1}{a\,p^*} + \frac{b\,p^*}{2}
\label{eq_effective}
\end{equation}
where $a =$ 4.03 $\pm$ 1.72 $\pm$ 0.06 GeV$^{-1}$ and $b =$ 1.29 $\pm$ 0.63 $\pm$ 0.67 GeV$^{-1}$. Removing the two-body phase space factor, given by $\frac{p^*}{m_{K\pi}}$, from LASS non-resonant amplitude, which is already included in Eq.~\ref{eq1}, we obtain the following parametrization for the non-resonant hadronic final state interaction: 
\begin{equation}
\mathcal{A}_{\textrm{NR}} = \frac{m_{K\pi}}{p^*}\,{\sin(\delta_{\textrm{LASS}}) 
e^{i\delta_{\textrm{LASS}}}}.
\label{eq3}
\end{equation}

The weak matrix element for the vector process, $\mathcal{M}_1$, and for the scalar process, $\mathcal{M}_0$, are written as a function of helicity amplitudes, $H_{i}$, derived in~\cite{schuler}. Neglecting the mass of the charged lepton the matrix elements are
\begin{multline}
\mathcal{M}_{1} = \sqrt{\qsq}~\left[ (1 + \cos\thl) \sin\thv e^{i\chi} H_{+}(\qsq,m_{K\pi}
) \right. \\
\left. - (1 - \cos\thl) \sin\thv e^{-i\chi} H_{-}(\qsq,m_{K\pi}) - 2 \sin\thl 
\cos\thv H_{0}(\qsq,m_{K\pi}) \right]
\label{eq4a}
\end{multline} 
and 
\begin{equation}
\mathcal{M}_{0} = - 2 \sqrt{\qsq}~\sin \thl ~H_{0}^s(\qsq,m_{K\pi}).
\label{eq4b}
\end{equation} 

The three form factors for the vector states and the one for the scalar states are written assuming the single pole dominance \emph{ansatz} given by:
\begin{equation}
f_{\textrm{ansatz}}(\qsq) = \frac{f(0)}{1 - q^2/M_{\textrm{pole}}^2}.
\end{equation}

The vector states use the nominal spectroscopic pole masses, $M_{A} =$ 2.5 GeV/$c^2$ and $M_{V} =$ 2.1 GeV/$c^2$, and the recent form factor measurements in Ref.~\cite{topher2}. The scalar states use $M_{V} =$ 2.1 GeV/$c^2$ and the respective zero recoil form factor is arbitrarily set to one since its value can always be absorbed in the amplitude parameter $a_{0,(R)}$. 

\mysection{Angular distribution study}

\begin{figure}[tbp]
\begin{center}
\includegraphics[width=4.0in]{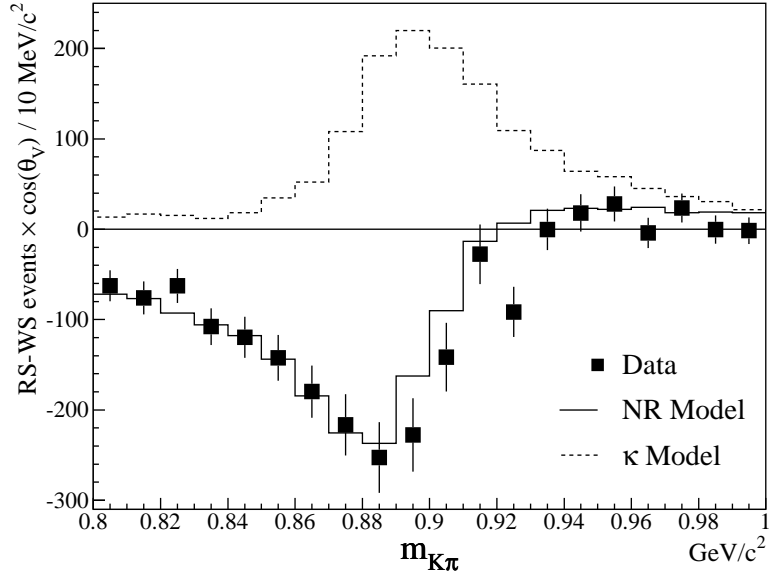}
\caption{
The background subtracted distribution of \mkpi{} weighted by $\cos{\thv{}}$. The data (squares) show good agreement with the LASS non-resonant parameterization (solid histogram) but not with a $\kappa$ model (dashed histogram)}
\label{fig_asymmetry}
\end{center}
\end{figure}

Next we discuss the angular distribution described by Eq.~\ref{eq1}. The $K\pi$ spectrum described by this equation includes the dominant contribution from the \kstar{} resonance, possible high mass contributions from the \klow{} and \khigh{} resonances, and low mass scalar components comprised of a non-resonant and a possible $\kappa$ contributions, both populating the region where relevant discrepancies were found. As discussed in~\cite{topher1,topher2}, the \mkpi{} distribution weighted by $\cos{\thv{}}$ provides information on the phase of the additional structure relative to that of the \kstar{}. It can be used to discriminate different combinations of low mass states, given the large difference between their phase shifts. \footnote{The expected \mkpi{} distribution weighted by $\cos{\thv{}}$ for a pure \Dkstarmunu{} decay would be nearly zero.} Figure~\ref{fig_asymmetry} compares the distribution obtained in the data with the predictions from the non-resonant and $\kappa$ models in the absence of additional phase shifts. 

Since a simulation using the LASS parametrization of the non-resonant contribution is sufficient to reproduce the data, we exclude a possible $\kappa$ contribution from further consideration.

Having excluded the $\kappa$, the most general differential decay rate for \Dkpimunu{} in \mkpi{} is given by Eq.~\ref{eq5a} where $\epsilon$ represents the detector acceptance and efficiency:

\begin{equation}
\frac{d\Gamma}{d m_{K\pi}} = \int \epsilon \left| \mathcal{M}_1 \mathcal{V} + \mathcal{M}_0 \mathcal{S} \right| ^2  ~\phi~ d\Omega
\label{eq5a}
\end{equation}

with vector and scalar amplitudes given by
\begin{gather}
\label{eqv} 
\mathcal{V} \equiv a_{K^{*}(892)^0} \mathcal{A}_{K^{*}(892)^0} + a_{K^{*}(1680)^0} \mathcal{A}_{K^{*}(1680)^0}\\[10pt]
\mathcal{S} \equiv a_{\textrm{NR}} \mathcal{A}_{\textrm{NR}} + a_{K^{*}_{0}(1430)^0} \mathcal{A}_{K^{*}_{0}(1430)^0}.
\label{eqS}
\end{gather}

The amplitude coefficients $a_{K^{*}(892)^{0}}$, $a_{K^{*}(1680)^{0}}$, $a_{\textrm{NR}}$, and $a_{K^{*}_{0}(1430)^{0}}$ are real, as required by Watson's Theorem~\cite{watson}. 

\mysection{The fit procedure}

\begin{figure}[tbp]
\begin{center}
\includegraphics[width=4.0in]{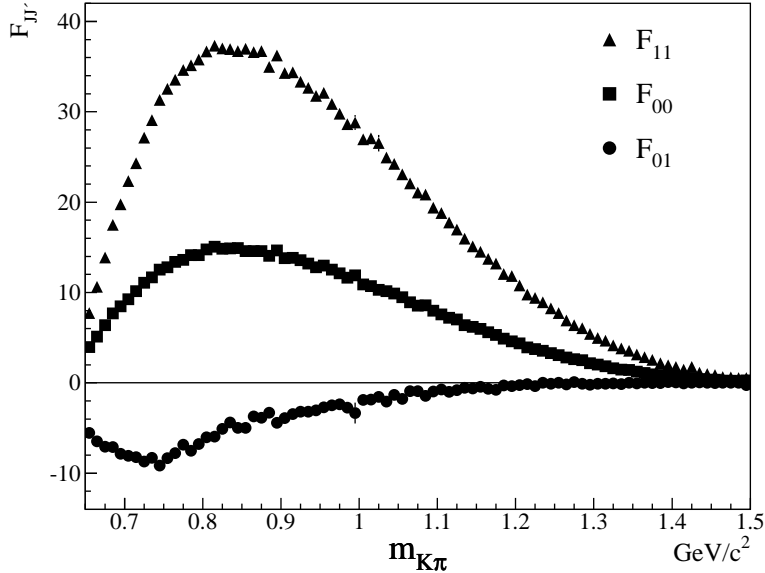}
\caption{Distributions of the relevant $F_{JJ^{'}}$ functions which represent the intensity modified by acceptance and efficiency and include the effects of phase space and the weak matrix elements.  The $F_{11}$, $F_{00}$, and $F_{01}$ distributions are shown as triangles, squares, and circles, respectively.
\label{fig_weight}}
\end{center}
\end{figure}

Equation~\ref{eq5a} can be conveniently factorized as:

\begin{equation}
\frac{d\Gamma}{d m_{K\pi}} = |\mathcal{V}|^2 F_{11} + |\mathcal{S}|^2 F_{00} + 2\,\Re (\mathcal{V^*~S}) F_{01} 
\label{eq5b}
\end{equation}

where $F_{JJ^{'}} \equiv \int \epsilon \mathcal{M}^*_{J} \mathcal{M}_{J^{'}}\,\phi\,d\Omega$, are real functions\footnote{Because of Eq.~\ref{eq4a}, all imaginary pieces of $\mathcal{M}^*_{J} \mathcal{M}_{J^{'}}$  will appear as sinusoidal functions of $\chi$. Hence any imaginary terms vanish when averaged over $\chi$ given our nearly uniform acceptance in this variable.} that depend only on \mkpi. The $F_{JJ^{'}}$ functions are computed from the \mkpi{} spectrum obtained from a complete simulation of \Dkpimunu{} events, generated according to phase space and weighted by $\mathcal{M}^*_{J} \mathcal{M}_{J^{'}}$ and thus represent the intensity modified by acceptance and efficiency. The three $F_{JJ^{'}}$ functions are shown in Figure~\ref{fig_weight}. The $|\mathcal{V}|^2$, $|\mathcal{S}|^2$, and  $\Re (\mathcal{V}^* \mathcal{S})$ functions depend on \mkpi{} as well as on all fit parameters. The cross-term, $2\,\Re (\mathcal{V^*~S}) F_{01}$, represents the interference between the vector and scalar contributions. 

The contribution from each decay mode, as well as the \kstar{} parameters, are obtained from an unbinned maximum-likelihood fit. We define the probability density function as the sum of the probability density for the signal, $\mathcal{L}_{S}$, and for the background, $\mathcal{L}_{B}$. The signal density is described by Eq.~\ref{eq5b}. The background density is given by the sum of charm and non-charm contributions. The relative contribution of the two background sources as well as the relative fraction of the background with respect to the selected \Dkpimunu{} sample, $f_B$, are fixed at the estimated values, described previously. We fit the data by minimizing the quantity $\omega$, 

\begin{equation}
\omega \equiv - 2 \ln{\sum_{\textrm{events}}\left[ (1 - f_{B}) \mathcal{L}_{S} + f_{B} \mathcal{L}_{B} \right]}
\label{likelihood}
\end{equation}  

The fit parameters are the magnitudes of each amplitude in the signal probability density function ($a_i$), the nominal mass and width of the \kstar{}, and the parameter $r_0$ of the Blatt-Weisskopf damping factor. The \kstar{} is taken as the reference amplitude ($a_{K^{*}(892)^0} = 1$). The parameters of all other resonances are fixed to the PDG values~\cite{pdg}. 
Decay fractions are obtained integrating each individual amplitude over the phase space and dividing by the integral over the phase space of the overall amplitude.

To account for momentum resolution effects on the \kstar{} parameters, we refit the data fixing all parameters except the \kstar{} width and use the probability density function, $\mathcal{L}^{G}$, given by Eq.~\ref{convolution}:

\begin{equation}
\mathcal{L}^{G}(m_{\textrm{K$\pi$}}) = \int \mathcal{L}(m_{\textrm{K$\pi$}}') \,G(m_{\textrm{K$\pi$}}'-m_{\textrm{K$\pi$}}, \sigma) \,d m_{\textrm{K$\pi$}}'.
\label{convolution}
\end{equation}

The new probability density function, $\mathcal{L}^{G}(m_{K\pi})$, represents the convolution of the data fit function with a Gaussian distribution, $G$, with $\sigma = 5.88~\textrm{MeV}/c^2$, value obtained from Monte Carlo simulation. The smearing due to momentum resolution increases the \kstar{} width by approximately $2~\textrm{MeV}/c^2$.

\mysection{The results}

Using the procedure described above, we fit the data assuming only a \Dkstarmunu{} process. The confidence level of this fit is 0.21\%, indicating the need for additional contributions in the decay.

\begin{figure}[tbp]
\begin{center}
\includegraphics[width=5.0in]{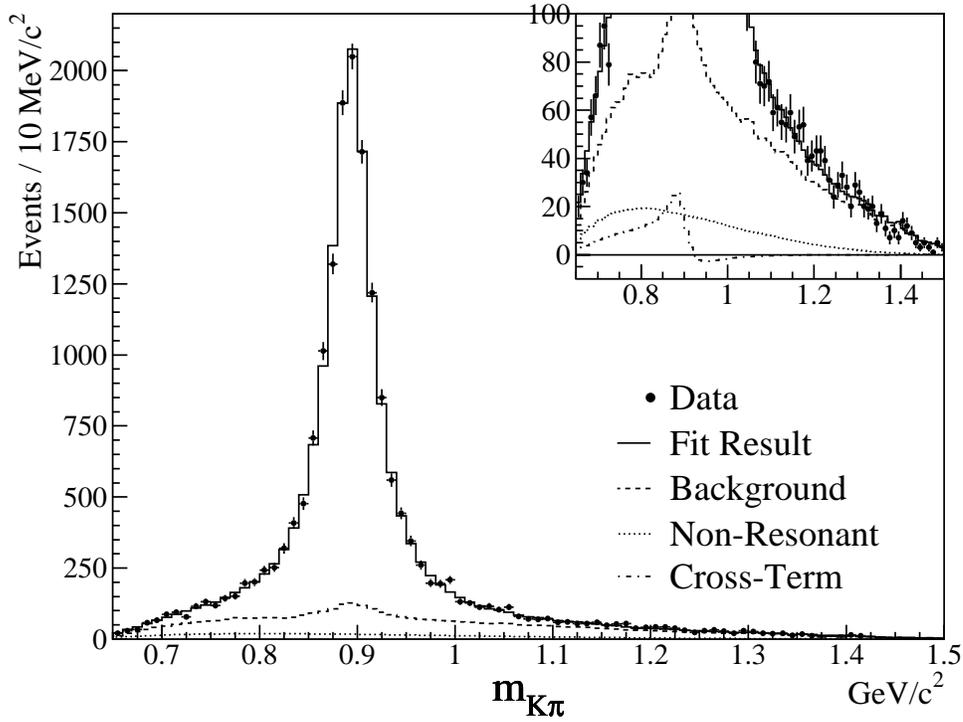}
\caption{Fit to the \mkpi{} data using the NR model. The error bars, the solid lines, the dashed lines, and the dotted lines correspond to the data, the model, the background contribution, and the scalar contribution, respectively. The upper right plot shows the same information and the cross-term (dot-dash line) with a limited y-axis to allow more detail to be seen. 
\label{fig_results}}
\end{center}
\end{figure}

The inclusion of a non-resonant scalar component, referred to as the \emph{NR model}, significantly improves the confidence level of the fit to 66\%.  We find $m_{K^{*}(892)^0} = 895.41 \pm 0.32$~MeV/c$^2$, $\Gamma_{K^{*}(892)^0} = 47.79 \pm 0.86$~MeV/c$^2$, $r_0 = 3.96 \pm 0.54$~GeV$^{-1}$, $a_{\textrm{NR}} = 0.327 \pm 0.024$, which correspond to a scalar fraction of $5.30 \pm 0.74$ \%. Figure~\ref{fig_results} illustrates the contribution of both the \Dkstarmunu{} and non-resonant s-wave process to the observed \mkpi{} spectrum. 

We also consider possible \Dkhighmunu{} and \Dklowmunu{} contributions to our model. Since the data is already well described by a model having only the \kstar{} and non-resonant components, we do not expect large contributions from these modes. Including both decays we find $m_{K^{*}(892)^0} = 895.0 \pm 1.1$~MeV/c$^2$, $\Gamma_{K^{*}(892)^0} = 47.63 \pm 0.91$~MeV/c$^2$, $r_0 = 5.7 \pm 4.8$~GeV$^{-1}$, $a_{\textrm{NR}} = 0.287 \pm 0.073$, $a_{{\overline{K}^{*}}(1680)^0} = -0.16 \pm 0.36$, and $a_{{\overline{K}^{*}_{0}}(1430)^0} = -0.048 \pm 0.19$. The \khigh{} and \klow{} amplitudes are consistent with zero and we find $\frac {\Gamma (D^+ \! \to \! \khighb \mu^+ \nu)} {\Gamma (D^+ \! \to \! K^- \pi^+ \mu^+ \nu)} < 4.0$\% and $\frac{\Gamma (D^+ \! \to \! \klowb \mu^+ \nu)} {\Gamma (D^+ \! \to \! K^- \pi^+ \mu^+ \nu)}<0.64$\% at 90\% CL. The upper limits are calculated using the method described in~\cite{lions} and assume $\textrm{BR}(\khighb \!\to\! K^-\pi^+)=0.258$ and $\textrm{BR}(\klowb \!\to\! K^-\pi^+)=0.62$~\cite{pdg}. When the \khigh{} is included, we observe a strong correlation between $r_0$ and $a_{{\overline{K}^{*}}(1680)^0}$, inflating the errors on both quantities. To study the statistical significance of these new amplitudes, we use a hypothesis test based on the maximum-likelihood ratio method~\cite{testhypothesis}. This method compares two hypotheses and points out unnecessary degrees of freedom. As a result, we obtain a confidence level of 80\% in favor of the simple NR model.

We consider several sources of systematic errors. These include variations of the fit conditions, {\em split sample} errors, and the uncertainty on the absolute mass scale of the experiment, relevant for the \kstar{} mass measurement. Twenty-seven variations of the fit procedure are considered. Starting from the final sample we adopt more stringent selection criteria changing the significance of the separation between secondary and primary vertices, the secondary vertex isolation requirement, and the cut on the muon confidence level. In addition, we vary the relative fractions of the different background components. We vary by $\pm 1\sigma$ the values of the parameters from the LASS effective range parametrization (Eq.~\ref{eq_effective}). We also include as a systematic error the difference between the FOCUS results, obtained from the NR model, and the results obtained from the model with high mass structures. Errors from this source are asymmetrical: we take the difference between the central and highest/lowest values of each fit parameter and scaled by 0.68 to obtain the contribution to the systematic error.

The {\em split sample} component takes into account systematic effects introduced by residual differences between data and Monte Carlo. This component is determined by splitting the data into five pairs of independent subsamples, according to the $D^{\pm}$ charge, data taking conditions, primary vertex multiplicity, muon momentum, and the momentum of the $K\pi$ system.  The treatment used for the split sample is known as Unconstrained Averaging, described in~\cite[page 14]{pdg}. 

The total systematic error is given by the sum in quadrature of the uncertainties from the independent sources. Table~\ref{tab_systematic} presents the results of the systematic uncertainty evaluation for the measurements.

\begin{table}[tbp]
\protect\caption{Estimated systematic uncertainty obtained for each component.}
\label{tab_systematic}
\begin{center}
\begin{tabular}{|c|c|c|c|}
\hline
\cline{2-4}
                                    & Cut and Model Vars. & Split Sample & Mass Scale \\ 
\cline{2-4}
$m_{K^{*}(892)^0}$ (MeV/$c^2$)      & ${}^{\,+\, 0.13}_{\,-\, 0.29}$  & $\pm 0.11$ & $\pm 0.30$ \\ 
$\Gamma_{K^{*}(892)^0}$ (MeV/$c^2$) & ${}^{\,+\, 0.81}_{\,-\, 0.18}$  & $\pm 1.05$ & --- \\
$r_0$ (GeV$^{-1}$)                  & ${}^{\,+\, 1.16}_{\,-\, 0.67}$  & $\pm 0.61$ & --- \\
Scalar fraction (\%)                & ${}^{\,+\, 0.95}_{\,-\, 0.92}$  & $\pm 0.28$ & --- \\
\hline 
\end{tabular}
\end{center}
\end{table}

\begin{figure}[tbp]
\begin{center}
\includegraphics[width=4.0in]{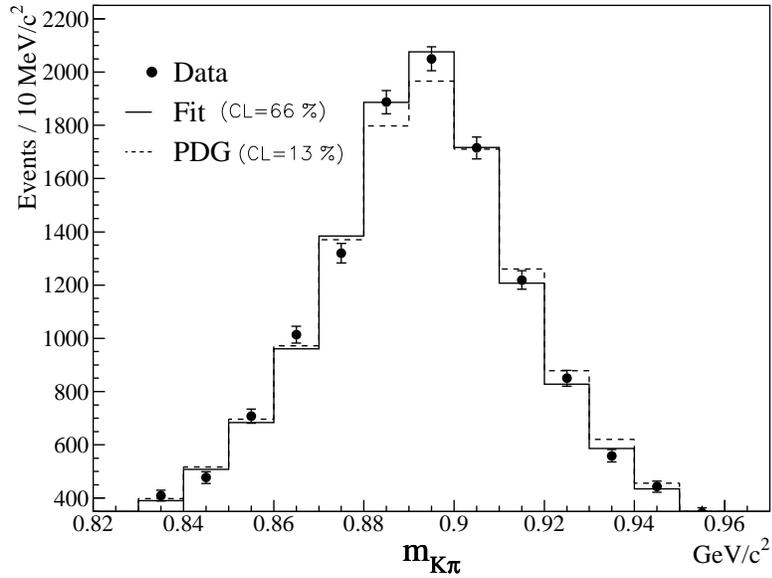}
\caption{The \mkpi{} spectrum in data (error bars) comparing to the NR model with free \kstar{} parameters (solid histogram) and the NR model with \kstar{} parameters fixed to the PDG~\cite{pdg} values (dashed histogram).
\label{fig_resolution}}
\end{center}
\end{figure}

Table~\ref{tab_results_1} summarizes the results obtained from the fits using the two models. The values of the fit parameters are compared to the world average values~\cite{e687,topher2,lass,pdg}. Our measurements of the \kstar{} mass and width are both more than $1\,\sigma$ below the PDG average values. Figure~\ref{fig_resolution} shows a comparison between our standard NR model with free \kstar{} parameters and a NR model with the mass and width of \kstar{} resonance fixed to the world average values~\cite{pdg}. With the inclusion of a non-resonant contribution, the value we obtain for the Blatt-Weisskopf parameter is consistent with LASS~\cite{lass}. The fraction of the scalar component is compatible with the value obtained previously in the analysis of the \costhv{} asymmetry~\cite{topher2}.

\begin{table}[tbp]

\protect\caption{Summary of results on \kstar{} parameters and contributions from non-\kstar{} sources in the decay \Dkpimunu{} obtained from the NR model. Fit result is compared to the current world averages and to the model with only \kstar. Limits on $\klowb$ and $\khighb$ contributions account for unseen decay modes.}

\label{tab_results_1}
\begin{center}
\begin{tabular}{cccc}
                                    & \kstar{} only & FOCUS result & Current values \\ 
\hline
$m_{K^{*}(892)^0}$ (MeV/$c^2$)      & 895.61 $\pm$ 0.32 & 895.41 $\pm$ 0.32$^{\,+\, 0.35}_{\,-\, 0.43}$ & 896.10 $\pm$ 0.27~\cite{pdg} \\ 
$\Gamma_{K^{*}(892)^0}$ (MeV/$c^2$) & 50.26 $\pm$ 0.81  & 47.79 $\pm$ 0.86$^{\,+\, 1.32}_{\,-\, 1.06}$  & 50.70 $\pm$ 0.60~\cite{pdg} \\
$r_0$ (GeV$^{-1}$)                  & 14.1 $\pm$ 5.7    & 3.96 $\pm$ 0.54$^{\,+\, 1.31}_{\,-\, 0.90}$   & 3.40 $\pm$ 0.67~\cite{lass} \\

\scriptsize{$\displaystyle \frac{\Gamma(D^+ \!\to\!K^-\pi^+\mu^+\nu)_\textrm{NR}}{\Gamma (D^+ \!\to\! K^- \pi^+ \mu^+ \nu)}$} (\%) & & 5.30 $\pm$ 0.74$^{\,+\, 0.99}_{\,-\, 0.96}$   & \andree{}{$\sim$ 5~\cite{topher2}}{8.3 $\pm$ 2.9~\cite{e687}} \\ 

\scriptsize{$\displaystyle \frac{\Gamma (D^+ \!\to\! \khighb \mu^+ \nu)}{\Gamma (D^+ \!\to\! K^- \pi^+ \mu^+ \nu)}$} & & $<$ $4.0\%$~@~90\%~CL & \\
\scriptsize{$\displaystyle \frac{\Gamma (D^+ \!\to\! \klowb \mu^+ \nu)}{\Gamma (D^+ \!\to\! K^- \pi^+ \mu^+ \nu)}$} & & $<$ $0.64\%$~@~90\%~CL & \\

Confidence level (\%)               & 0.21              & 66.0 &  \\ 
\hline
\end{tabular}
\end{center}
\end{table}

\mysection{Summary and Discussion}
       
In conclusion we have measured the \kstar{} parameters using a large sample of \Dkstarmunu{} signal events over a wide mass range. The absence of high mass resonances as well as the small background contribution provides a unique environment to study the \kstar{} mass and width. The \kstar{} mass and width measurements are stable with respect to model variation. Our measurements of the mass and width are more than $1\,\sigma$ below the present world average value. We obtain a Blatt-Weisskopf parameter consistent with the value obtained by LASS~\cite{lass}. We also limit possible additional $K\pi$ resonances present in \Dkpimunu{} semileptonic decays. Our angular distribution is consistent with the effective-range scalar non-resonant phase shift obtained by LASS~\cite{lass} as expected by Watson's Theorem given the absence of other final state interactions. 

\mysection{Acknowledgments}

We wish to acknowledge the assistance of the staffs of Fermi National Accelerator Laboratory, the INFN of Italy, the Universidade Federal do Rio de Janeiro, the Funda\c c\~ao de Amparo \`a Pesquisa do Estado do Rio de Janeiro and the physics departments of the collaborating institutions. This research was supported in part by the U.~S. National Science Foundation, the U.~S. Department of Energy, the Italian Instituto Nazionale di Fisica Nucleare and Ministero dell'Universit\`a e della Ricerca Scientifica e Tecnol\'ogica, the Brazilian Conselho Nacional de Desenvolvimento Cient\'{\i}fico e Tecnol\'ogico, CONACyT-M\'exico, the Korean Ministry of Education, and the Korea Research Foundation.


\begin{thebibliography}{99}


\bibitem{bigi} 
I. I. Bigi and A. I. Sanda, 
{\it CP Violation}, 
Cambridge University Press, (1999) page 55.

\bibitem{watson} 
K. M. Watson, 
\prd{88}, 5 1163 (1952).

\bibitem{kstardominance} 
M. Adamovich \etal~(WA82 Collaboration), 
\plb{268}, 142 (1991).

\bibitem{e687} 
P.~L.~Frabetti \etal~(E687 Collaboration), 
\plb{307}, 262 (1993).

\bibitem{blatt} 
J. M. Blatt and V. F. Weisskopf, 
{\it Theoretical Nuclear Physics}, Wiley, New York (1952).

\bibitem{topher1} 
J. M. Link \etal~(FOCUS Collaboration), 
\plb{535}, 43 (2002).

\bibitem{topher2} 
J. M. Link \etal~(FOCUS Collaboration), 
\plb{544}, 89 (2002).

\bibitem{dkpipi} 
E. M. Aitala \etal~(E791 Collaboration), 
\prl{89}, 121801 (2002). 

\bibitem{isgur} 
D. Scora and N. Isgur, 
\prd{52}, 2783 (1995).

\bibitem{dosch} 
P. Ball, V. M. Braun and H. G. Dosch, 
\plb{273}, 316 (1991).

\bibitem{marina} 
H. G. Dosch, E. M. Ferreira, F. S. Navarra and M. Nielsen, 
\prd{65}, 114002 (2002).

\bibitem{beam}
P.~L.~Frabetti \etal,
Nucl.\ Instrum.\ Meth.\ A {\bf 329}, 62 (1993).

\bibitem{detector1}
P.~L.~Frabetti \etal~(E687 Collaboration),
Nucl.\ Instrum.\ Meth.\ A {\bf 320}, 519 (1992).

\bibitem{detector2}
J.~M.~Link \etal~(FOCUS Collaboration),
Nucl.\ Instrum.\ Meth.\ A {\bf 484}, 270 (2002).

\bibitem{detector3}
J.~M.~Link \etal~(FOCUS Collaboration),
Nucl.\ Instrum.\ Meth.\ A {\bf 516}, 364 (2002).


\bibitem{pythia} 
T. Sj\"ostrand, 
Computer Physics Commun. 82 (1994) 74.           

\bibitem{lass} 
D. Aston \etal~(LASS Collaboration), 
\npb{296}, 493 (1988).


\bibitem{schuler} 
J.~G.~Korner and G.~A.~Schuler, 
\zpc{46}, 93 (1990).

\bibitem{pdg} 
S. Eidelman \etal, 
\plb{592}, 1 (2004). 

\bibitem{lions} L.~Lyons, 
{\it Statistics for nuclear and particle physicists}, 
Cambridge University Press, page 78.

\bibitem{testhypothesis} F. James, 
{\it Determining the Statistical Significance of Experimental Results}, 
Technical Report DD/81/02 CERN (1981).

\end{thebibliography}
\end{document}